# Phase synchronization recovery in energy-compressed LWFA electron beams for free-electron lasers via undulator tapering

Shan-You Teng[1,2], Wai-Keung Lau[2], Shih-Hung Chen[1*], Wei-Yuan Chiang[2†]

[1]Department of Physics, National Central University, Taoyuan 320317, Taiwan

[2]National Synchrotron Radiation Research Center, Hsinchu 300092, Taiwan

**Abstract**

Laser wakefield accelerators (LWFAs) are attractive compact drivers for free-electron lasers (FELs) because they can generate femtosecond electron beams with high peak current over centimeter-scale acceleration distances. However, their relatively large energy spread remains a major obstacle to high-gain FEL operation. Although bunch energy compression can reduce the slice energy spread to a level suitable for FEL amplification, it also introduces a strong energy chirp. The energy chirp detunes the FEL resonance along the planar undulator, causing phase slippage between the electrons and the radiation field, reduced bunching efficiency, and degraded radiation power and spectral quality. Here we investigate a longitudinally tapered undulator for compensating the chirp-induced resonance mismatch in a self-amplified spontaneous-emission (SASE) FEL driven by an energy-compressed LWFA beam. Using three-dimensional unaveraged simulations, we show that an optimized taper profile restores electron-radiation phase synchronization and significantly improves both the saturation power and the spectral properties relative to the untapered case. We also assess the sensitivity of the scheme to shot-to-shot beam-energy fluctuations characteristic of LWFA operation. Our results show that undulator tapering is an effective method for mitigating chirp-induced performance degradation in compact plasma-based FELs.

[*]chensh@ncu.edu.tw
[†]chiang.wy@nsrrc.org.tw

## I. INTRODUCTION

Free-electron lasers (FELs) are recognized as fourth-generation light sources capable of producing radiation with high brilliance, full transverse coherence, and ultrashort pulse duration. As accelerator-based sources, they offer broad wavelength tunability from the terahertz to the hard x-ray regime [1-7]. However, FEL performance depends strongly on the quality of the driving electron beam. In conventional radio-frequency (RF) linac-based facilities, the electron energies required for extreme-ultraviolet (EUV) and x-ray generation typically necessitate beamlines extending over tens to hundreds of meters, leading to large and costly infrastructures. In recent years, laser wakefield accelerators (LWFAs) have emerged as promising alternatives because their GV/m-scale accelerating gradients enable the production of GeV-class electron beams within centimeter-scale plasmas [8–11]. Their femtosecond bunch duration and sub-micrometer transverse emittance at bunch charges of a few tens of pC make them attractive candidates for compact FEL drivers. However, their relatively large divergence, substantial energy spread, and strong shot-to-shot fluctuations remain major obstacles to high-gain FEL operation.

For beam transport from an LWFA to an FEL, careful control is required to limit the rapid beam-size growth caused by the large divergence at the plasma exit and to preserve the transverse emittance along the transport line, so that the beam quality required in the FEL interaction region can be maintained. At the same time, the beam energy spread must be reduced because excessive energy spread suppresses FEL gain and limits coherent amplification. Several techniques have been proposed to address this challenge [12]. Plasma-based methods, such as density tailoring and down-ramp injection, can reduce the intrinsic chirp during acceleration [13]. Beamline-based methods include transverse-gradient undulators (TGUs) with dedicated dispersive transport, which allow electrons with different energies to remain resonant with the radiation field through a transverse variation of the undulator field [14]. Another option is a magnetic bunch energy-compression section based on dipoles, which can reduce the slice energy spread to a level below the FEL parameter.

Remarkable progress has been made toward proof-of-principle demonstrations of LWFA-driven FELs. Although LWFA beams typically have energy spreads at the few-percent level, a major breakthrough was reported in 2021 by the Shanghai Institute of Optics and Fine Mechanics (SIOM), where plasma density profiling was used to control the beam energy chirp and produce a 490 MeV electron beam with 0.5% energy spread, sub mm·mrad emittance, and 30pC bunch charge. This beam successfully drove a self-amplified spontaneous-emission (SASE) FEL at 27 nm with radiation power gain approaching two orders of magnitude [15,16]. With more refined plasma density tailoring and reduced bunch charge, the same approach has also produced electron beams with per-mille and even sub-per-mille energy spread [17]. Rather than reducing the energy spread directly at the source, the COXINEL project at SOLEIL demonstrated seeded FEL operation at 270 nm by applying energy compression to an LWFA beam with percent-level energy spread [18]. More recently, stable operation of a 420 nm LWFA-driven SASE FEL with radiation power gain exceeding three orders of magnitude was demonstrated at LBNL using a similar energy-compression strategy [19]. Together, these pioneering results demonstrate the growing feasibility of compact high-gain FELs driven by LWFAs.

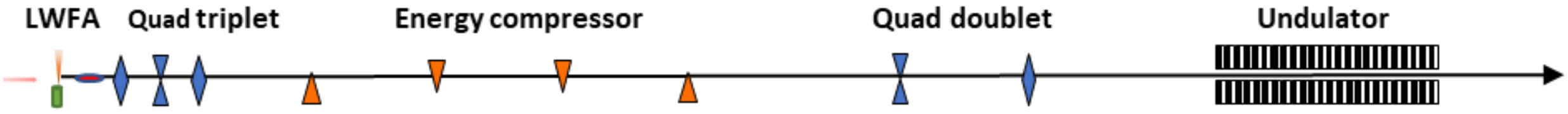

Figure 1: Schematic layout of a LWFA-driven FEL beamline with energy compression

However, magnetic bunch energy compression inevitably introduces a energy chirp through the energy dependence of the particle path length, and this chirp degrades FEL performance. Plasma dechirpers provide one possible remedy [20,21] by using wakefield-induced fields to compensate the correlated energy variation along the bunch. Their experimental implementation, however, remains challenging because of the intrinsic shot-to-shot fluctuations of LWFA beams and the stringent synchronization required between multiple plasma stages. A conceptually simpler alternative is to employ a longitudinally tapered undulator, as discussed in Refs. [22–24]. In such a scheme, the taper compensates for the chirp-induced resonance mismatch and helps maintain efficient beam-radiation coupling over the undulator length. This concept is particularly relevant for LWFA-driven FELs, in which the correlated chirp may be introduced intentionally by magnetic bunch energy compression used to reduce the slice energy spread. The design must therefore balance the reduction of slice energy spread against the preservation of sufficient peak current, while using the undulator taper to restore resonance and recover FEL gain. In this paper, we propose an LWFA-based FEL scheme at NCU, shown in Fig. 1, for vacuum-ultraviolet amplification at 66.5 nm. The scheme combines a four-dipole energy compressor with a longitudinally tapered undulator. Start-to-end simulations based on beam parameters expected from ongoing LWFA experiments show that this approach restores electron-radiation phase synchronization and significantly improves FEL performance, providing a practical route toward compact plasma-based coherent light sources.

This paper is organized as follows. In Sec. II, we describe the proposed scheme and derive the taper gradient required for chirp compensation. Section III presents the simulation setup for the NCU FEL beamline. In Sec. IV, we analyze the FEL performance improvement obtained with the tapered undulator and evaluate the sensitivity of the scheme to beam-energy jitter. Section V summarizes the conclusions.

## II. PHASE-SYNCHRONIZATION RECOVERY BY UNDULATOR TAPERING

### A. Chirp-induced resonance detuning and taper condition

We consider an ultrashort electron bunch produced by an LWFA and transported through a beamline containing a magnetic energy compressor. The purpose of the compressor is to reduce the slice energy spread to a level suitable for FEL amplification. Because the intrinsic bunch length of an LWFA beam is typically much shorter than the bunch-length contribution arising from energy-spread-induced longitudinal dispersion in the compressor, the bunch length $\sigma_z'$ after an energy compressor may be estimated as

$$\sigma_z' \sim | R_{56} | \sigma_\delta, \tag{1}$$

where $| R_{56} |$ is the longitudinal dispersion of the compressor and $\sigma_\delta$ is the rms energy spread of the beam. After compression, the electron bunch typically has an approximately linear energy chirp, namely a linear correlation between electron energy and longitudinal position. The chirp parameter is defined as $h \equiv \frac{d\delta}{dz}$, then we have

$$h \sim \frac{1}{| R_{56} |}, \tag{2}$$

where $\delta \equiv \frac{\Delta\gamma}{\gamma_0}$. To clarify how a linear energy chirp affects the electron dynamics in the undulator, we examine the energy-dependent longitudinal path-length difference. For an electron with a relative energy deviation $d\delta \equiv \frac{d\gamma}{\gamma_0}$, the corresponding longitudinal shift can be expressed through the longitudinal dispersion of the undulator [25].

$$\frac{dz}{d\delta} \approx \frac{L_u}{{\gamma_0}^2}\left(1 + \frac{K^2}{2}\right), \tag{3}$$

where $\gamma_0$ is the average beam energy, $L_u$ is the propagation distance in the undulator and K is the undulator parameter. In the FEL, the characteristic spacing between adjacent microbunches is approximately equal to the resonant wavelength. By taking integration over the energy difference between adjacent microbunches, i.e., $\Delta\delta = h\lambda_r$, where $\lambda_\mathrm{r} = \frac{\lambda_u}{2{\gamma_0}^2}\left(1 + \frac{K^2}{2}\right)$ is the resonant wavelength [26], we have

$$\Delta z \approx \frac{L_u}{{\gamma_0}^2}\left(1 + \frac{K^2}{2}\right)\lambda_r/|R_{56}|, \tag{4}$$

Thus, the initial microbunch spacing corresponding to the resonant wavelength $\lambda_r$ is gradually altered during propagation through the undulator. This accumulated longitudinal shift introduces a phase mismatch between the electron density modulation and the co-propagating radiation field, as shown in Fig. 2. For a beam with large energy chirp, the bunch length effectively elongates in the undulator due to longitudinal dispersion, which interrupts the continuous interaction between the radiation field and the electrons [25]. As shown in Fig. 2, an energy-chirped electron beam experiences unequal longitudinal dispersion along the undulator, leading to a progressive evolution of the microbunch spacing and phase detuning with respect to the radiation field.

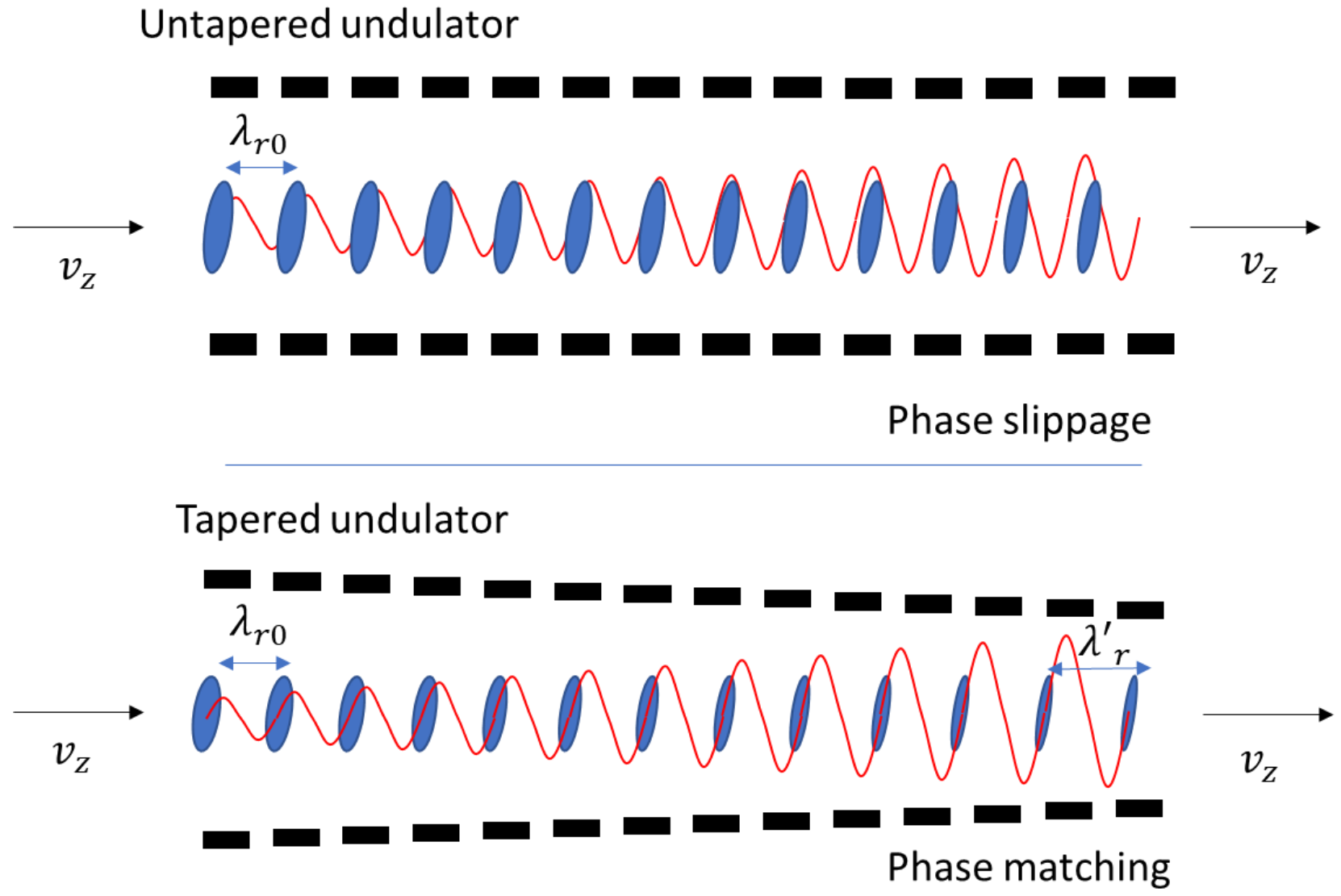

Figure 2: Conceptual illustration of chirped-beam FEL interaction and taper compensation.

To restore synchronism between the electron beam and radiation fields, the undulator resonance condition must be dynamically adjusted so that the radiation wavelength follows the evolving microbunch spacing. By introducing a longitudinal taper, corresponding to a gradual variation of the undulator parameter K, the resonant wavelength can be tuned to compensate for the chirp-induced longitudinal evolution. The modified resonance condition can then be expressed as

$$\lambda_r' = \lambda_r + \Delta z = \frac{\lambda_u}{2{\gamma_0}^2}\left(1 + \frac{K'^2}{2}\right), \tag{5}$$

where $\lambda_u$ is the undulator period length and $K'$ is the effective undulator parameter introduced to incorporate the longitudinal shift correction. Substituting Eq. (4) into Eq. (5) yields

$$\lambda_r + \frac{L_u}{{\gamma_0}^2}\left(1 + \frac{K_0^2}{2}\right)\lambda_r/R_{56} = \frac{\lambda_u}{2{\gamma_0}^2}\left(1 + \frac{K'^2}{2}\right) \tag{6}$$

Solving Eq. (6) for the effective undulator parameter, the required modification to the undulator strength can be explicitly expressed as

$$K'^2 = K_0^2 + \frac{2L_u}{\gamma_0^2}\left(1+\frac{K_0^2}{2}\right)^2 /|R_{56}| \tag{7}$$

which shows that the required increase in $K'^2$ is directly determined by the magnitude of the chirp-induced longitudinal shift and is proportional to the undulator length. Since the correction term is relatively smaller than the nominal $K_0$, Eq. (7) can be approximately expressed as

$$K' \sim K_0 + \frac{L_u}{\gamma_0^2}\left(1+\frac{K_0^2}{2}\right)^2 /|R_{56}| \tag{8}$$

For practical implementation, it is more convenient to express the variation of the undulator parameter in terms of its longitudinal gradient. By differentiating the effective undulator parameter with respect to the longitudinal coordinate, the required taper rate can be written as

$$\frac{dK}{dz} = \frac{1}{{\gamma_0}^2}\left(1+\frac{K_0^2}{2}\right)^2 /|R_{56}|. \tag{9}$$

This relation indicates that once the energy compressor's longitudinal dispersion $|R_{56}|$ is determined, the required undulator taper rate $\mathrm{dK/dz}$ can be uniquely specified to compensate for the induced energy chirp and preserve the phase synchronization along the undulator.

Furthermore, by substituting the linear energy chirp for $|R_{56}|$ and using the relation $dz = -cdt$ to express the evolution in terms of time, the taper condition can be rewritten as

$$\frac{dK}{dz} = \frac{1}{{\gamma_0}^3}\left(1+\frac{K_0^2}{2}\right)^2 * \frac{\mathrm{d}\gamma}{dz} = -\frac{1}{{\gamma_0}^3} * \frac{d\gamma}{cdt}\left(1+\frac{K_0^2}{2}\right)^2, \tag{10}$$

Interestingly, this relation is consistent with the compensation condition previously obtained through numerical simulations of attoseconds pulses generation in Ref [23]. Our results provide additional physical insight into the mechanism of chirp compensation that the undulator taper adjusts the resonant wavelength along the undulator so that it follows the evolution of the microbunch spacing in the chirped electron beam.

In the present work, however, the chirp is not externally imposed for pulse shaping but arises naturally from the magnetic energy-compression stage used to prepare the LWFA beam for FEL amplification. The tapered undulator therefore serves as a practical means of recovering synchronism between the emitted radiation and the microbunch, while preserving the benefit of reduced slice energy spread.

The present analysis relies on two approximations. First, the final bunch length is estimated using Eq. (1), neglecting the initial bunch length. This approximation is valid when the energy spread dominates. Second, the expansion of Eq. (9) assumes the correction term is small. These approximations may lose accuracy for very small energy spread or relatively low energy. However, for typical LWFA beams, which are characterized by high energy, large energy spread, and ultrashort bunch length, both assumptions are generally well satisfied.

## B. Beamline configuration and simulation strategy

The proposed FEL beamline is designed to combine slice-energy-spread reduction through magnetic compression with energy-chirp compensation using a tapered undulator. As shown in Fig. 1, the beamline consists of three main sections: a transport line from the LWFA source, a four-dipole magnetic compressor, and a planar undulator section with an adjustable longitudinal taper. The transport line is used to capture and match the LWFA beam into the compression and undulator system while minimizing emittance growth and excessive beam-size increase caused by the large divergence at the plasma exit. The compressor is then used to reduce the slice energy spread to a level compatible with FEL amplification, while the undulator section converts the conditioned beam into coherent VUV radiation.

In the study, the beam parameters are chosen based on those expected from ongoing LWFA experiments at NCU. The reference beam energy is selected to satisfy the FEL resonance condition at a target wavelength of 66.5 nm. The initial bunch duration, peak current, normalized emittance, and energy spread are taken from LWFA beam estimates and are used as input parameters for the downstream beamline simulations. Particular attention is given to the trade-off introduced by magnetic compression. Stronger compression can effectively reduce the slice energy spread, but it also introduces an energy chirp and may reduce the peak current available for FEL gain if the longitudinal phase space becomes significantly distorted. Therefore, the compressor strength is chosen to balance slice-energy-spread reduction against current degradation.

The undulator section is modeled as a planar radiator with period $\lambda_u$ and reference undulator parameter $K_0$. The taper is introduced as a longitudinal variation of the undulator parameter, $K(z)$, starting from the value required to satisfy the nominal resonance condition for the reference beam energy. The analytical estimate in Eq. (9), derived in Sec. II A is used as the initial guideline for setting the taper rate, while the final taper profile is determined through numerical optimization. In this way, the taper is treated as a control parameter for recovering phase synchronization between the chirped electron beam and the radiation field after magnetic compression.

To evaluate the FEL performance, we perform three-dimensional unaveraged simulations using PUFFIN [27], which self-consistently models the interaction between the electron beam and the radiation field without relying on the slowly varying envelope approximation. This approach is particularly suitable for LWFA-driven FELs, where the beam duration is ultrashort, the energy chirp can be significant, and the longitudinal phase-space distribution may deviate substantially from the assumptions of conventional time-independent or averaged FEL models. In practice, Eq. (9) provides only a first-order estimate of the required taper because the FEL interaction is also affected by three-dimensional effects, finite bunch duration, and the detailed longitudinal phase-space distribution of the compressed LWFA beam. In the simulations presented below, this analytical estimate is therefore used as the initial guideline for determining the taper profile, while the final taper strength is optimized numerically to maximize FEL performance.

## III. SIMULATION OF LWFA BASED FEL BEAMLINE

To evaluate the performance of the proposed LWFA-driven FEL at NCU, we use a start-to-end simulation framework combining IMPACT-T [28,29] and PUFFIN. Beam transport from the LWFA source through the magnetic energy compressor and into the undulator is modeled with IMPACT-T, including three-dimensional space-charge effects and coherent synchrotron radiation (CSR). The beamline considered here consists of a four-dipole magnetic energy compressor followed by a planar undulator for vacuum-ultraviolet (VUV) radiation generation. The simulation parameters adopted in this study are summarized in Table I. The reference LWFA electron beam has an energy of 250 *MeV*, a bunch charge of 50 *pC*, and a bunch duration of 2.12 *fs*. The normalized emittance and intrinsic relative energy spread are assumed to be 0.5 *mm·mrad* and 1%, respectively, while the initial transverse beam size and divergence are taken as 0.72 *μm* and 1.4 *mrad*. These parameters are consistent with representative beam qualities expected from ongoing LWFA experiments and are suitable for assessing the feasibility of compact FEL amplification in the VUV regime.

After transport through the compressor section, the electron beam is injected into a planar undulator with period length $\lambda_u = 20$ mm, undulator parameter $K_u = 1.09$, and total number of periods $N_u = 400$. These parameters correspond to a resonant radiation wavelength of 66.5 *nm* for the reference beam energy. A longitudinal taper is applied to compensate for the chirp-induced resonance detuning generated during magnetic compression. In this study, the taper gradient $\frac{dK}{dz}$ is varied from 0.009 to 0.038 $m^{-1}$, depends on the compression strength. Because the taper is obtained by numerical optimization guided by the first-order estimate described in Sec. II A, it should be interpreted as an effective taper parameter for the present beamline configuration.

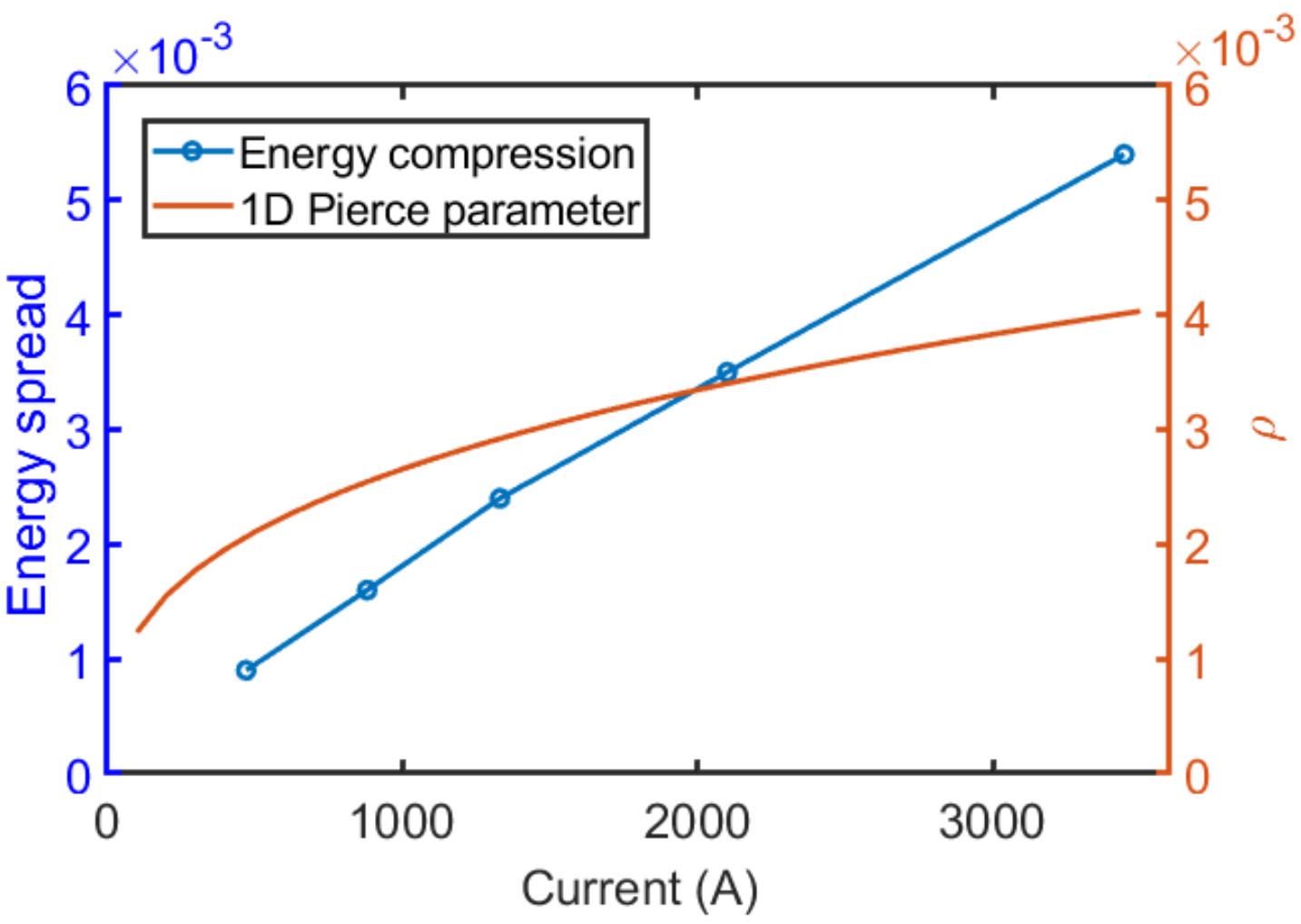

FIG. 3. Dependence of the slice energy spread and the 1D Pierce parameter ρ on the beam current for different energy-compression settings (10, 15, 20, 25, 35mrad)

The magnetic compressor provides a tunable longitudinal dispersion by varying the dipole bending angle. Each dipole has a length of $0.1m$, and the drift length between adjacent dipoles is $0.4m$. The bending angle is scanned from 10 to $35mrad$ to examine different compression strengths. Varying the bending angle changes the longitudinal dispersion $|R_{56}|$, thereby modifying the bunch length, slice energy spread, peak current, and energy chirp.

The optimal compression strength is determined by balancing the reduction of slice energy spread against peak-current degradation and enhanced energy chirp. Figure 3 shows the slice energy spread and the corresponding one-dimensional Pierce parameter ρ, calculated from the compressed beam current, as functions of the dipole bending angle. The results indicate that, over an appropriate range of bending angles, the slice energy spread can be reduced below the Pierce parameter, satisfying the high-gain FEL requirement $\sigma_\delta < \rho$ [26]. This condition defines the compression regime in which FEL amplification can be achieved despite the relatively large initial energy spread of the LWFA beam.

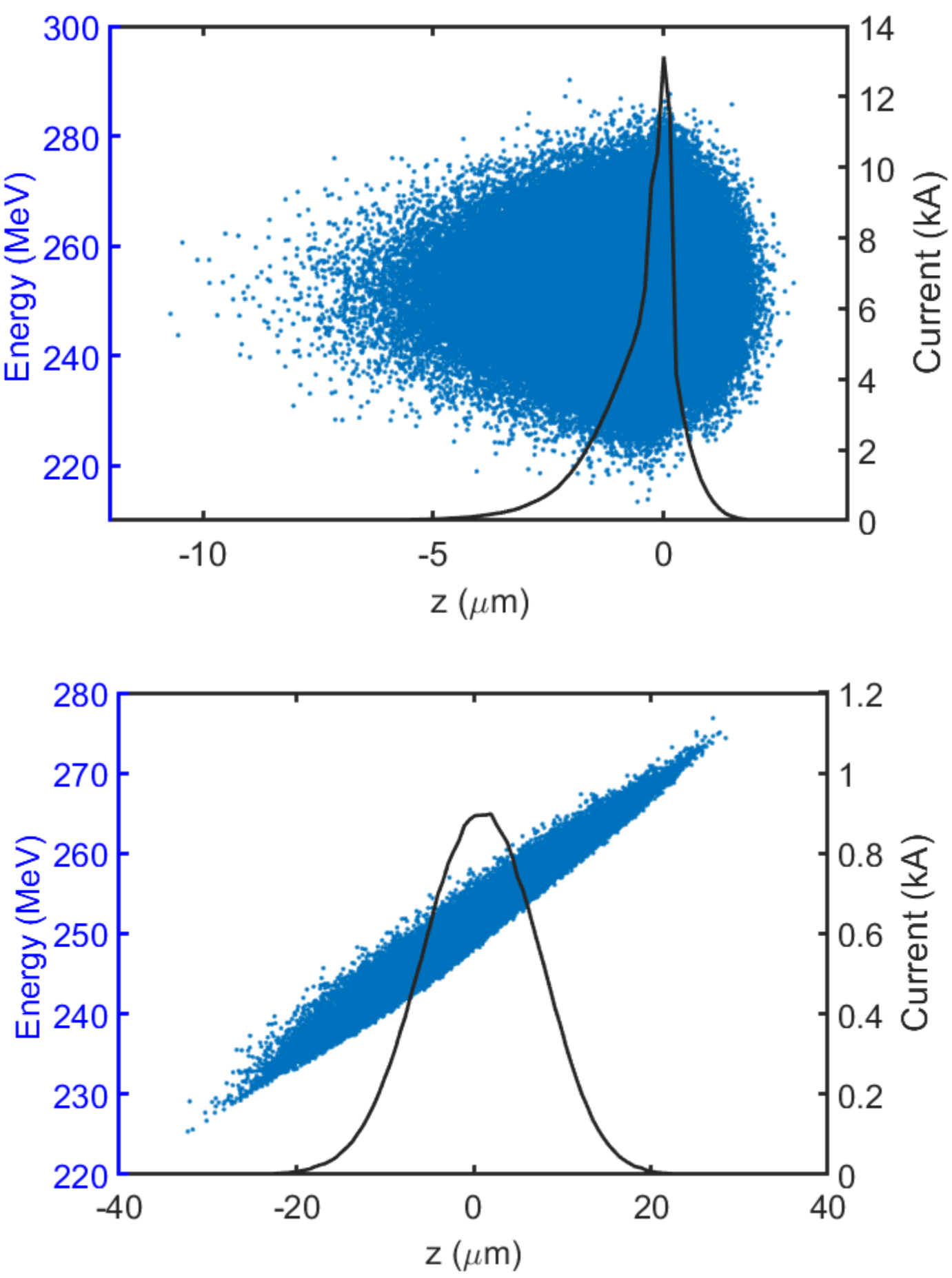

FIG. 4. Longitudinal phase-space distribution and current profile of the LWFA beam (a) before and (b) after the energy compressor with 25mrad bending angle. The energy compression process reduces the slice energy spread and produces an energy chirp.

Figure 4 shows the longitudinal phase-space distribution and current profile of the LWFA beam before and after magnetic compression for a dipole bending angle of 25 mrad. Before compression, the beam has an ultrashort bunch length of approximately 2 $\mu m$ , a relatively large intrinsic energy spread of 1% and a high peak current of about 13 kA. After passing through the compressor, the bunch is stretched to approximately 20 $\mu m$ and develops an approximately linear energy chirp. At the same time, the slice energy spread is reduced to about 0.15%, while the peak current decreases to about 1 kA. These results illustrate the central trade-off introduced by magnetic compression: the slice energy spread is substantially reduced, but at the cost of bunch lengthening, enhanced correlated chirp, and reduced peak current.

The compressed beam distribution is then used as the input to the PUFFIN simulations, and the corresponding FEL performance for different compression strengths is analyzed in the following section.

Table I. Simulation parameters of the NCU VUV FEL

| | **Parameter** | **Value** |
|---|---|---|
| **Electron beam** | Energy | 250 MeV |
| | Bunch charge | 50 pC |
| | Bunch duration | 2.12 fs |
| | Normalized emittance | 0.5 mm mrad |
| | Initial beam size | 0.72 μm |
| | Initial beam divergence | 1.4 mrad |
| | Energy spread | 1 % |
| **Undulator** | Period length $\lambda_u$ | 20 mm |
| | Number of periods $N_u$ | 400 |
| | Undulator parameter $K_u$ | 1.09 |
| | Taper gradient *dK/dz* | 0.009-0.038 /m |
| **Energy compressor** | Bending Angle | 10-35mrad |
| | Dipole length | 0.1m |
| | Drift length | 0.4m |

## IV. FEL SIMULATION RESULTS

### A. SASE simulation

The FEL amplification process is simulated with PUFFIN using the energy-compressed beam distribution as input. Figure 5 shows the evolution of the radiation peak power along the undulator for different energy compression strengths, together with the corresponding reference case without longitudinal tapering. In the absence of tapering, the energy chirp causes continuous phase slippage between the electrons and the radiation field, which weakens the beam-radiation coupling and leads to early saturation at only a few megawatts [27]. When the undulator taper is applied according to Eq. (9), this chirp-induced phase slippage is compensated along the undulator, allowing sustained energy exchange and continued power growth up to the gigawatt level at saturation. The final output power is thereby enhanced by more than two orders of magnitude compared with the untapered case.

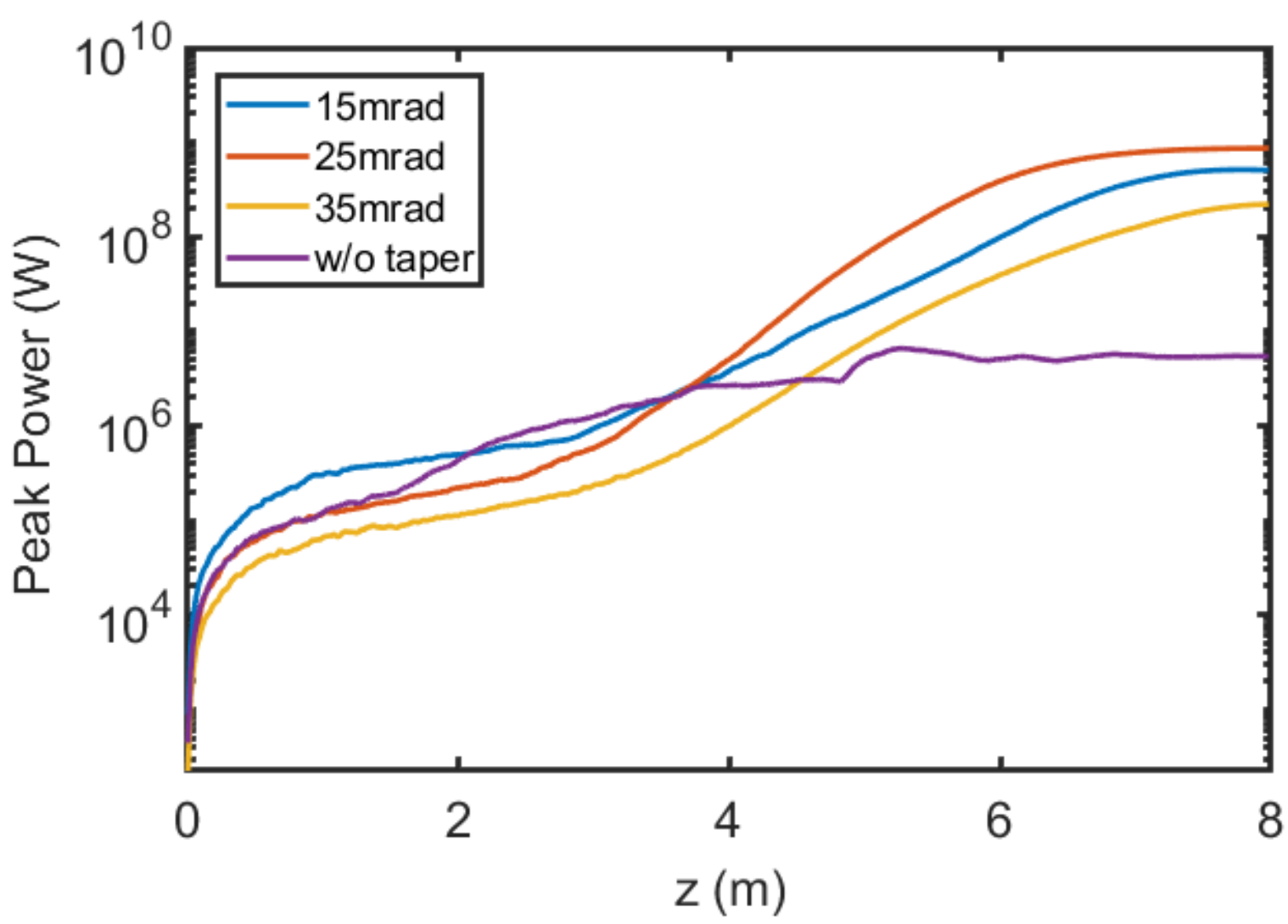

FIG. 5. Evolution of the FEL radiation peak power along the undulator for cases with different energy compression strength and without longitudinal tapering.

The corresponding bunching-factor evolution for the optimized case is shown in Fig. 6. Without tapering, the bunching factor reaches its maximum value prematurely, indicating that microbunching growth is interrupted by the progressive loss of phase synchronism. In contrast, the tapered case exhibits stronger and more sustained bunching growth throughout the interaction region. This behaviour is consistent with the enhanced power evolution shown in Fig. 5 and confirms that the tapered undulator maintains the coupling between the electron beam and the radiation field over a longer distance. In other words, the taper effectively compensates the chirp-induced resonance mismatch and restores the continuous microbunching process required for efficient FEL amplification. Taken together, Figs. 5 and 6 show that the principal role of the taper is to preserve beam-radiation synchronism along the undulator, thereby enabling continued bunching growth and the associated power enhancement.

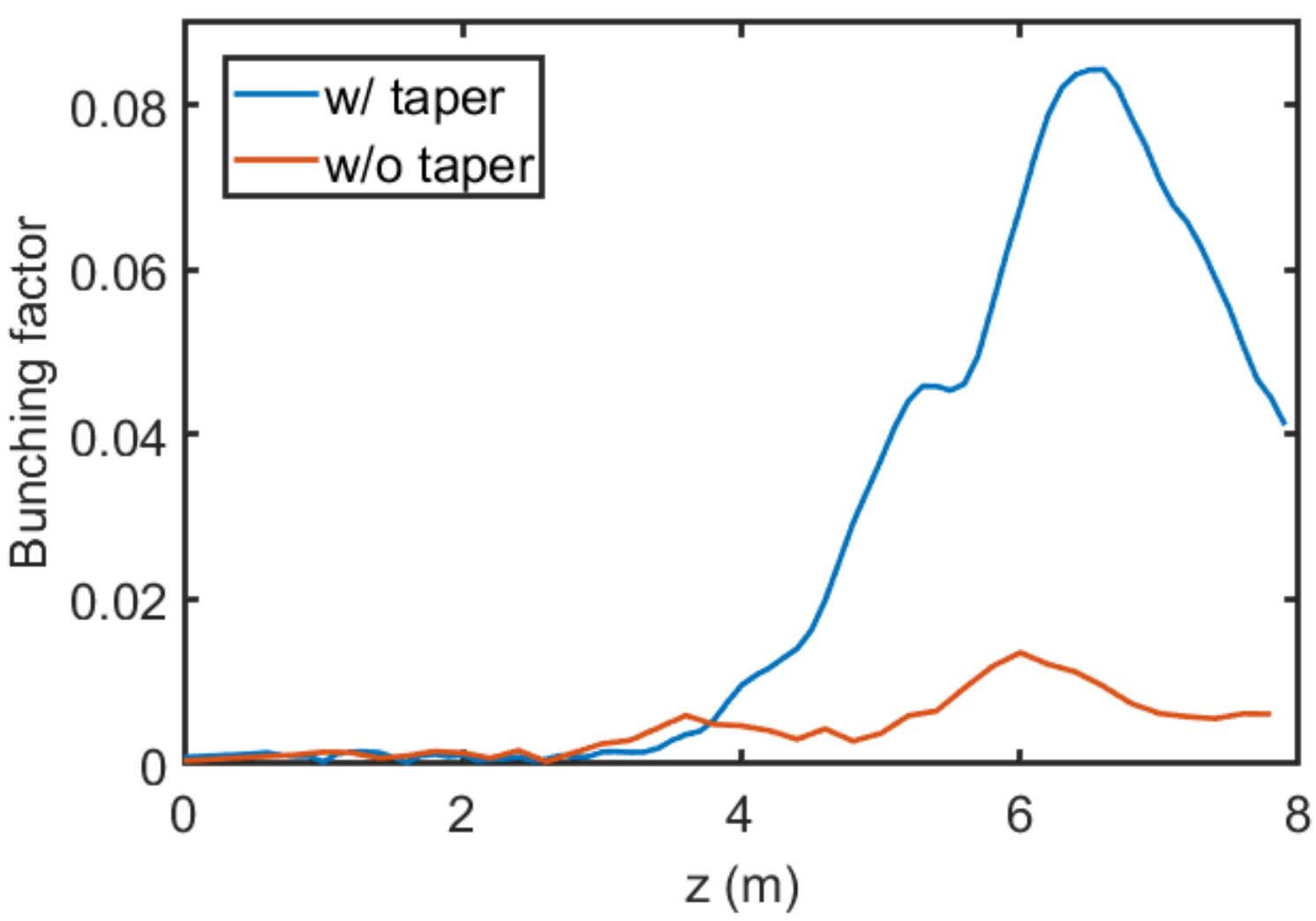

FIG. 6. Evolution of the bunching factor at resonant wavelength along the undulator for the optimized case with longitudinal tapering and without longitudinal tapering.

Figure 7 shows the output spectra for three representative compression strengths corresponding to dipole bending angles of 15, 25, and 35 mrad. The results illustrate how the energy chirp affects the spectral structure of the emitted radiation. For the 25 mrad case, the spectrum remains single-peaked but has both a redshift of the central wavelength and noticeable spectral broadening. This behavior results from the combined effects of bunch lengthening, energy chirp, and the chirp-compensating undulator taper. As the chirped beam propagates through the tapered undulator, the resonance condition is continuously adjusted so that the radiation wavelength follows the evolving microbunch spacing. Consequently, the effective lasing wavelength shifts to longer wavelengths, and the output spectrum becomes broader.

In contrast, the 15 *mrad* case has a clear double-peaked spectrum. This behavior is attributed to the stronger residual energy chirp in this case. Because the local resonant wavelength varies more significantly along the bunch, radiation emitted from different longitudinal slices is amplified at slightly different wavelengths. When the variation in resonant wavelength across the effective lasing region becomes sufficiently large, the amplified radiation separates into two nearby spectral components, producing a split spectrum.

A different behavior is observed for the 35 *mrad* case, in which the dominant radiation wavelength is shorter than the nominal resonant wavelength corresponding to the reference beam energy. According to the FEL resonance condition, a shorter wavelength corresponds to a higher electron energy, indicating that the lasing process is dominated by higher-energy slices of the bunch rather than by the central-energy slice. This can be understood from the reduced peak current in the 35 mrad compression case. The lower peak current leads to a longer FEL gain length, so that significant power growth occurs only after a longer propagation distance in the undulator. Under these conditions, amplification is dominated by the portion of the bunch that remains most effective for beam-radiation interaction, which in the present case corresponds to a higher-energy region near the leading portion of the bunch. In addition, because the energy chirp is weaker in this case, Eq. (9) predicts a smaller taper gradient. The resulting variation of the resonant wavelength along the undulator is therefore more limited, and the spectral broadening is less pronounced than in the 15 and 25mrad cases.

These results indicate that the spectral behavior shown in Fig. 7 is governed primarily by the strength and distribution of the energy chirp. The taper derived from Eq. (9) plays the role of maintaining phase synchronization and sustaining microbunching growth, but it does not by itself determine whether the spectrum remains single-peaked or develops multiple components. Instead, the detailed spectral structure is determined by the interplay among the energy chirp, the current profile, and the part of the bunch that dominates the FEL interaction.

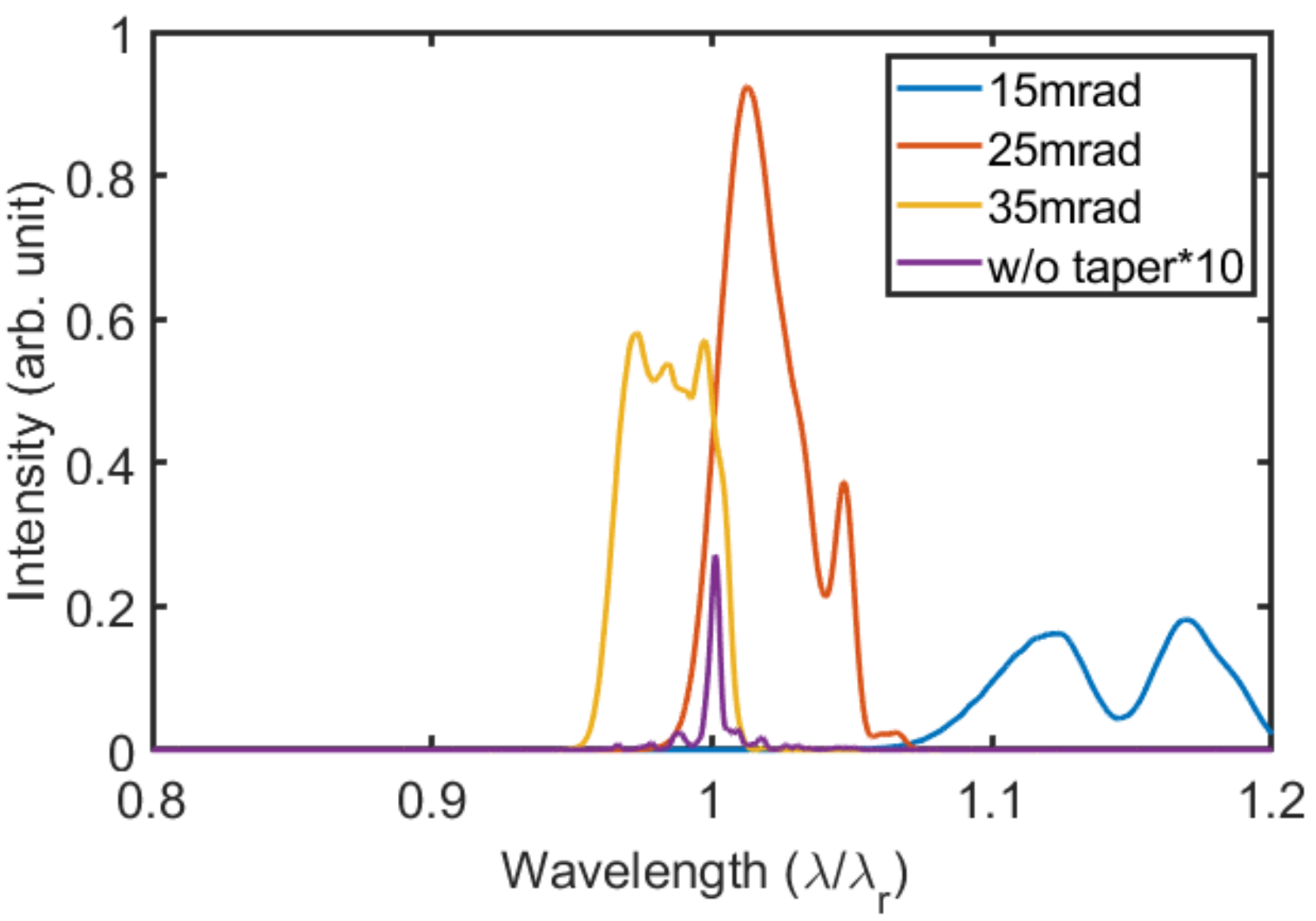

FIG. 7. Normalized output spectra at saturation for different energy-compression bending angles (15, 25, and 35 mrad). The 25 mrad case shows highest intensity and spectral localization, indicating optimal energy compression setting.

## B. Sensitivity studies

To further assess the robustness of the proposed scheme, sensitivity studies were conducted by varying both the electron-beam energy spread and the central beam energy, which are susceptible to shot-to-shot fluctuations in LWFA operation. Figure 8 shows the dependence of the saturation power on the electron-beam energy spread under two tapering conditions. In the first case, the undulator taper is fixed according to Eq. (9). In the second case, the taper strength is numerically optimized for each energy-spread condition.

The saturation power shown in Fig. 8 is evaluated by averaging the FEL power over the saturated region of the undulator rather than using the absolute peak value, in order to reduce the influence of post-saturation oscillations.

The results show that the fixed taper predicted by Eq. (9) agrees well with the numerically optimized taper over most of the energy-spread range. This indicates that the taper condition derived from the chirp-synchronization analysis provides a reliable first-order estimate for preserving efficient FEL amplification. In both cases, the saturation power generally decreases as the energy spread increases because the larger slice energy spread weakens the FEL gain and reduces the microbunching efficiency.

A weak change in the trend can also be observed around the intermediate energy-spread region in both curves. This behaviour is likely associated with the interplay between the peak current obtained after energy compression and the effective portion of the bunch that participates in the FEL interaction. Although a smaller energy spread leads to a higher peak current, the corresponding reduction in the effective lasing slices of the bunch may partially offset the gain enhancement. Nevertheless, this effect remains secondary compared with the overall degradation of FEL performance at larger energy spread.

A more noticeable deviation between the fixed-taper and optimized-taper cases appears in the low-energy-spread regime. This behavior originates from the approximation used in deriving Eq. (9), where the bunch length after the energy compressor is estimated using Eq. (1) under the assumption that the contribution from the initial bunch length is negligible. When the energy spread becomes sufficiently small, however, the compressed bunch length is no longer dominated by the energy-spread-induced compression term, and the initial bunch length contribution becomes non-negligible. Consequently, the actual bunch length after compression becomes longer than that predicted by Eq. (1), leading to a weaker energy chirp than expected. Under this condition, a smaller undulator taper is sufficient to maintain phase synchronization between the electron beam and the radiation field. As a result, the numerically optimized taper strength becomes systematically smaller than the fixed taper predicted by Eq. (9) in the low-energy-spread limit. Nevertheless, the agreement between the two tapering approaches remains reasonably good over the practical operating range relevant to the proposed LWFA-driven FEL.

.

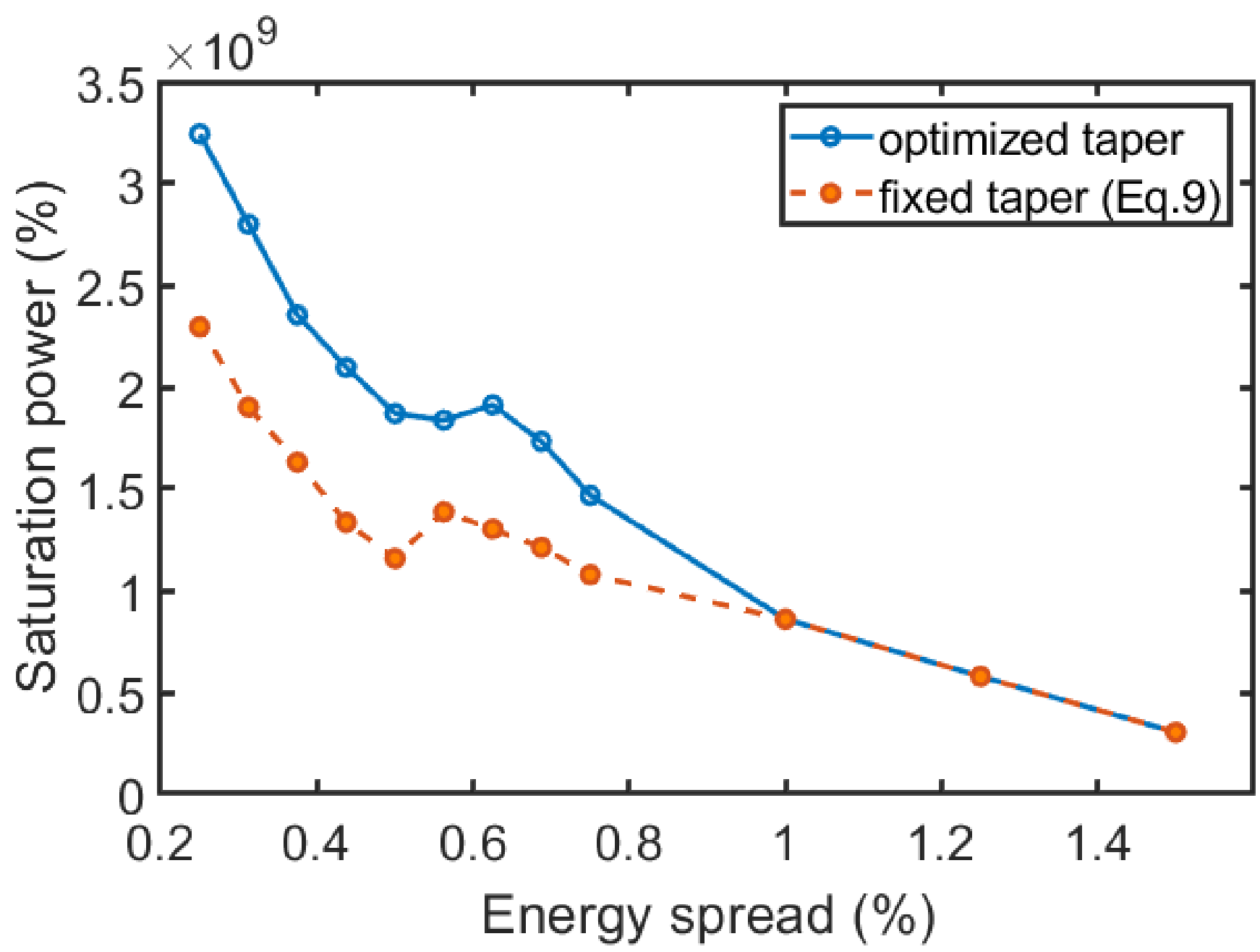

FIG. 8. Saturation power as a function of electron beam energy spread for the fixed taper predicted by Eq. (9) and the numerically optimized taper.

The deviation between the optimized taper and the taper predicted by Eq. (9) is quantified in Fig. 9, which shows the ratio of the numerically optimized taper strength to the analytical taper strength obtained from Eq. (9). The ratio approaches unity as the energy spread increases, indicating that the analytical approximation becomes increasingly accurate in the regime when the compressed bunch length is dominated by the energy-spread-induced compression term.

In contrast, the ratio decreases systematically in the low-energy-spread regime, where the optimized taper becomes significantly smaller than the analytical prediction. This behavior further confirms that the assumption used in Eq. (1), namely that the initial bunch length can be neglected after compression, gradually breaks down when the energy spread becomes small. In this regime, the energy chirp after compression is weaker than expected, and therefore a smaller taper strength is sufficient to maintain synchronization between the electron beam and the radiation field. Nevertheless, even in the low-energy-spread regime, the deviation remains within approximately 20–25%, demonstrating that Eq. (9) still provides a useful first-order estimate for practical taper design.

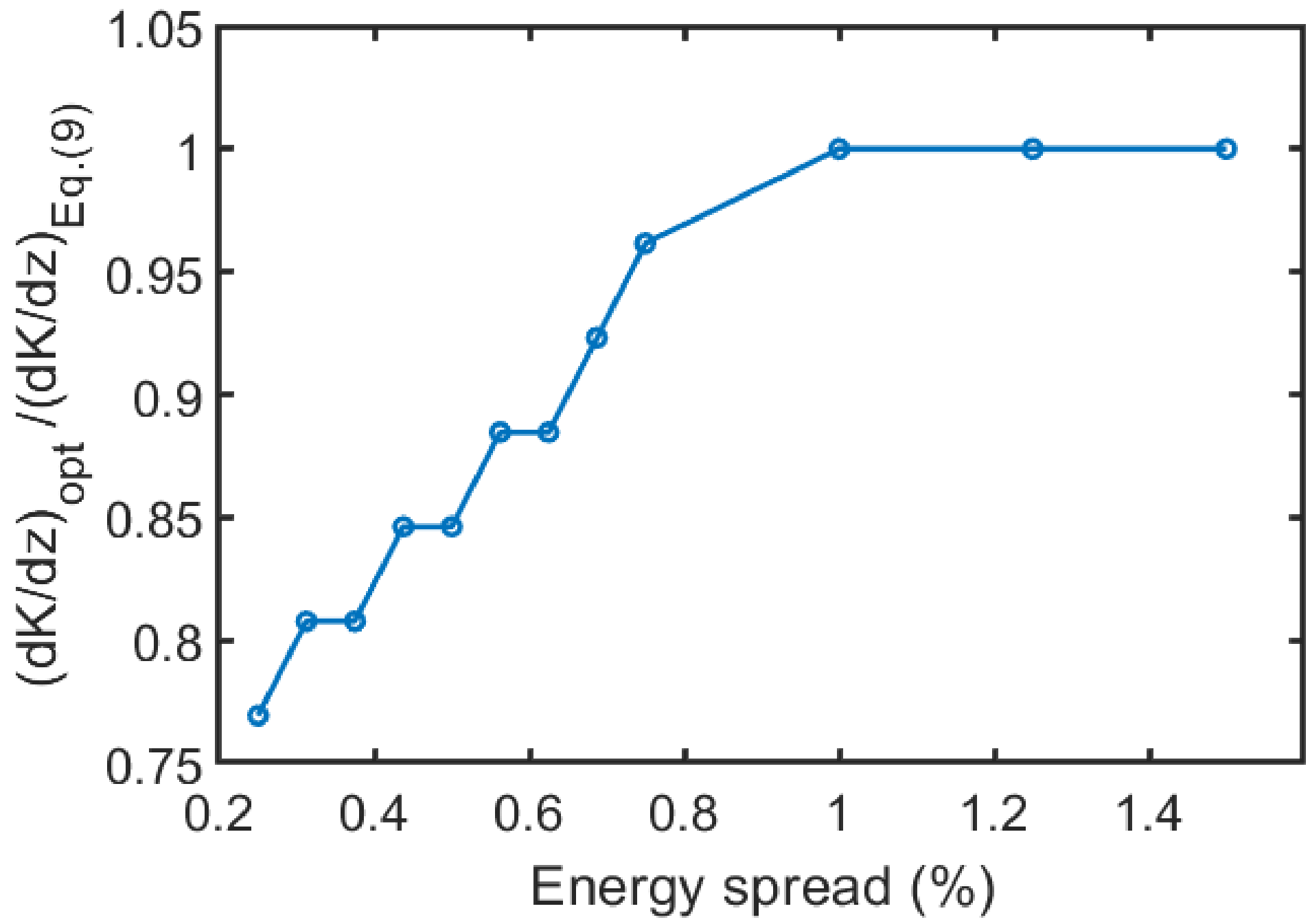

FUG. 9. Ratio of the numerically optimized taper strength to the taper strength predicted by Eq. (9) as a function of the electron-beam energy spread.

Next, FIG. 10 shows the dependence of the saturation power on the central beam-energy variation for two different beamline configurations, both using taper strengths calculated from Eq. (9). In the first case, the beamline optics are re-optimized for each beam-energy shift. Specifically, the quadrupole strengths were scanned to rematch the electron beam into the undulator for each central energy condition. The optimization was performed by minimizing the betatron oscillation amplitude and maintaining matched beam-size evolution inside the undulator, thereby maximizing the FEL interaction efficiency. In the second case, the quadrupole settings are fixed at the nominal 250 MeV condition.

For the re-optimized beamline, the saturation power remains relatively stable within a finite range around the nominal beam energy, indicating that the tapered FEL configuration is tolerant to central-energy variations when proper beam matching is maintained. In contrast, when the quadrupole settings are kept fixed, the saturation power decreases rapidly as the beam energy deviates from the nominal value. This degradation is mainly caused by mismatch of the beam optics, which increases the betatron oscillation amplitude in the undulator and thereby suppresses the FEL gain.

These results show that, unlike the energy-spread case, where taper matching is the dominant factor, the sensitivity to central beam-energy variation is governed primarily by the transport optics and matching conditions. In practice, however, it is difficult to maintain fully optimized beam matching for every LWFA shot. Shot-to-shot fluctuations of the central beam energy therefore become an important factor affecting FEL performance, and their impact must be carefully considered in the design and operation of LWFA-driven FEL beamlines.

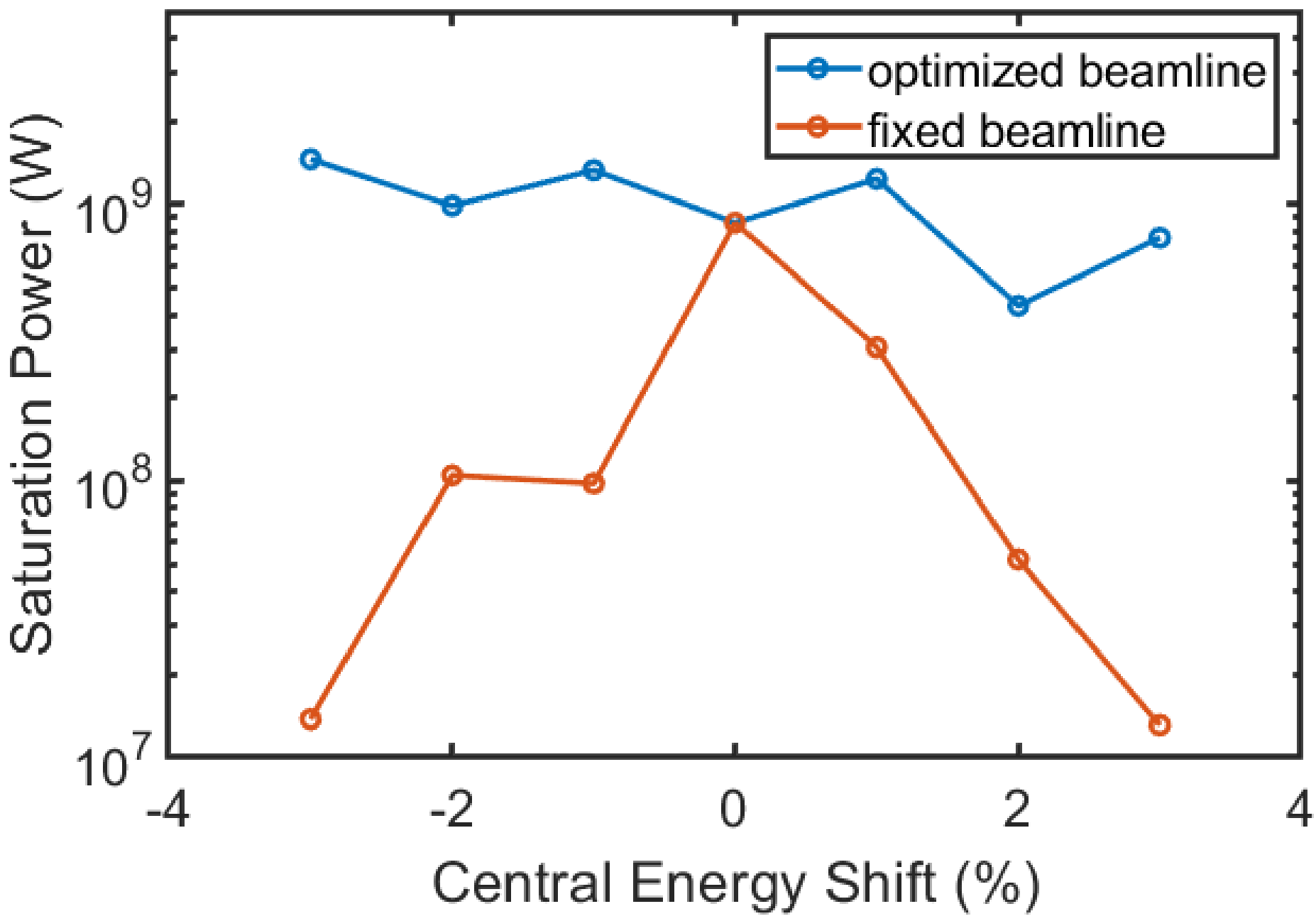

FIG. 10. Saturation power as a function of the central beam-energy shift for different beamline configurations. The optimized case corresponds to re-matched quadrupole settings for each energy shift, whereas the fixed case uses the quadrupole settings optimized for a 250MeV electron beam.

**CONCLUSION**

In this study, we analyzed a compact LWFA-driven FEL scheme using theoretical modeling and start-to-end simulations. A four-dipole magnetic compressor was employed to reduce the slice energy spread below the Pierce parameter while preserving sufficient peak current for FEL gain. In addition, a matching condition between the compressor-induced linear energy chirp and the undulator taper gradient was derived by requiring phase synchronism along the undulator. For each compression setting, the taper gradient was estimated from the derived expression and then applied in the FEL simulations.

Three-dimensional PUFFIN simulations show that the corresponding taper compensates the chirp-induced phase mismatch, restores phase synchronism, and enables sustained power growth to the gigawatt level using state-of-the-art LWFA electron-beam parameters. The spectral analysis further shows that the magnitude of the energy chirp controls the variation of the resonant wavelength and, consequently, the output spectral properties. An optimal compression strength is identified, for which the radiation spectrum remains localized near the design wavelength.

Sensitivity studies were also performed to assess the robustness of the proposed scheme. The results indicate that the taper condition derived from Eq. (9) remains effective over a broad range of energy spreads, providing robust compensation of the chirp-induced phase mismatch and preserving high output power. A noticeable deviation occurs only in the low-energy-spread regime, where the initial bunch length becomes non-negligible and a smaller optimal taper is required. The sensitivity to central-energy shifts further highlights the importance of reducing shot-to-shot beam-energy fluctuations for stable FEL operation.

These results demonstrate that chirp-taper matching provides a practical route toward high-brightness FEL amplification driven by LWFA electron beams. In this scheme, undulator tapering is not used merely as a post-saturation optimization technique, but serves as a mechanism for recovering the phase synchronism degraded by magnetic compression. The proposed approach therefore offers a promising path toward compact plasma-based coherent light sources.

## ACKNOWLEDGEMENTS

The authors would like to express our appreciation to Dr. Ji Qiang of LBNL for his guidance to use IMPAC-T code for this study and Dr. Shao-Wei Chou for intensive discussions on the design study of a compact VUV FEL. The successful completion of this research was made possible by the academic resources and advanced research infrastructure provided by the National Center for High-Performance Computing, National Institutes of Applied Research (NIAR), Taiwan. We gratefully acknowledge their invaluable support. This work was partially supported by the National Science and Technology Council (NSTC), Taiwan, under Grant number NSTC 113-2112-M-213-024, NSTC 113-2119-M-001-007, NSTC 113-2112-M-008-010, and 114-2112-M-008 -023 -MY2.